\documentclass[12pt]{iopart}
\usepackage{iopams}
\usepackage{graphicx}

\begin{document}
\title{Design of nanophotonic circuits for autonomous subsystem quantum error correction}

\author{J Kerckhoff$^\ast$, D S Pavlichin, H Chalabi and H Mabuchi}

\address{Edward L.\ Ginzton Laboratory, Stanford University, Stanford, California 94305, USA}
\ead{$^\ast${\tt jkerc@stanford.edu}}

\begin{abstract}
We reapply our approach to designing nanophotonic quantum memories to formulate an optical network that autonomously protects a single logical qubit against arbitrary single-qubit errors.  Emulating the 9 qubit Bacon-Shor subsystem code, the network replaces the traditionally discrete syndrome measurement and correction steps by continuous, time-independent optical interactions and coherent feedback of unitarily processed optical fields. 
\end{abstract}
\submitto{\NJP}

\maketitle

Traditional approaches to designing quantum memories presume a complex {\it classical} apparatus operating in parallel with the memory to detect and correct errors as they arise \cite{Gott09,QCQI}.  Typically, the mechanics of this essential half of the memory system are hardly considered at all, taking a back seat in the analysis to the qubits in which the information is stored.  This assumption of a perfect, classical overseer glosses over the inherent technological mismatch of typically fast, nanoscale and cold quantum systems with slow, mesoscopic, and/or hot classical ones.  The extreme performance demands for quantum information processing should motivate solutions that weigh classical and quantum resources wholistically and utilize system models as close to physical mechanics as possible. 

In \cite{Kerc10} we described an approach to designing protected quantum memory devices naturally suited to nanophotonic implementations that is technologically homogeneous and requires no ``off-chip'' oversight, simply ``power'' in the form of cw laser inputs.  The particular design example in \cite{Kerc10} emulates the bit- or phase-flip quantum error correcting (QEC) code \cite{Gott09,QCQI} with a single logical qubit stored in the collective state of three multi-level ``atoms,'' each strongly coupled to three different optical resonators.  Two more cavity quantum electrodynamical (cQED) systems serve as the binary ``controllers'' \cite{Mabu09b} for these qubits.  These five cQED devices are connected by a network of single-mode waveguides and beamsplitters that when appropriately powered by cw laser inputs continually and simultaneously cause the internal states of the controllers to reflect the joint parities of the qubits and corrective feedback on the qubits to be implemented according to the controller states.  Essentially a bit-/phase-flip QEC code, the design in \cite{Kerc10} can only protect the logical qubit against either single-qubit bit-flip or single-qubit phase-flip errors, depending on the configuration.  To emphasize how straightforwardly the approach scales to more powerful QEC codes, in this article we describe, model and simulate an autonomous nanophotonic network that emulates the 9 qubit Bacon-Shor subsystem code \cite{Baco06,Alif07}, the smallest of a class of naturally fault-tolerant QEC codes capable of protecting a single logical qubit from arbitrary single-qubit errors (see \ref{BS9} for a description of this code).     

While these networks are conceived of intuitively, our analysis arises from a mathematical framework of open quantum optical systems that takes quantum field theory as its physical basis \cite{Gardin92, OSQO} but most closely resembles a quantum generalization of electrical circuit theory \cite{Gough09,Nurd09a}.  Although rigorous stochastic dynamical modeling is often unfamiliar to physicists, this formalism \cite{HP84} is very physically intuitive once internalized and sufficiently flexible to describe the continuous-time dynamics of most systems foreseeable in quantum optical networks.  In our case, cQED devices are modeled as distinct Hamiltonian systems that couple weakly to free, bosonic fields that scatter off of each device in series and in parallel along paths set by the single-mode waveguides linking the network devices \cite{Gough09}.  As our network operates ``autonomously,'' without any user monitoring necessary, the equation of motion that describes the closed loop time-evolution of the devices (i.e. the dynamics after tracing over the field degrees of freedom) is a deterministic {\it master equation} \cite{OSQO,Kerc10}.  Moreover, as all the Hamiltonian and device-waveguide couplings are constant in time, the network is {\it stationary}.  Much like an electronic operational amplifier with a feedback impedance network, together the cQED memory and controllers represent an integrated, self-stabilizing system that simply requires DC ``power'' to function.

\begin{figure}[tb!]
\begin{center}\includegraphics[width=1\textwidth]{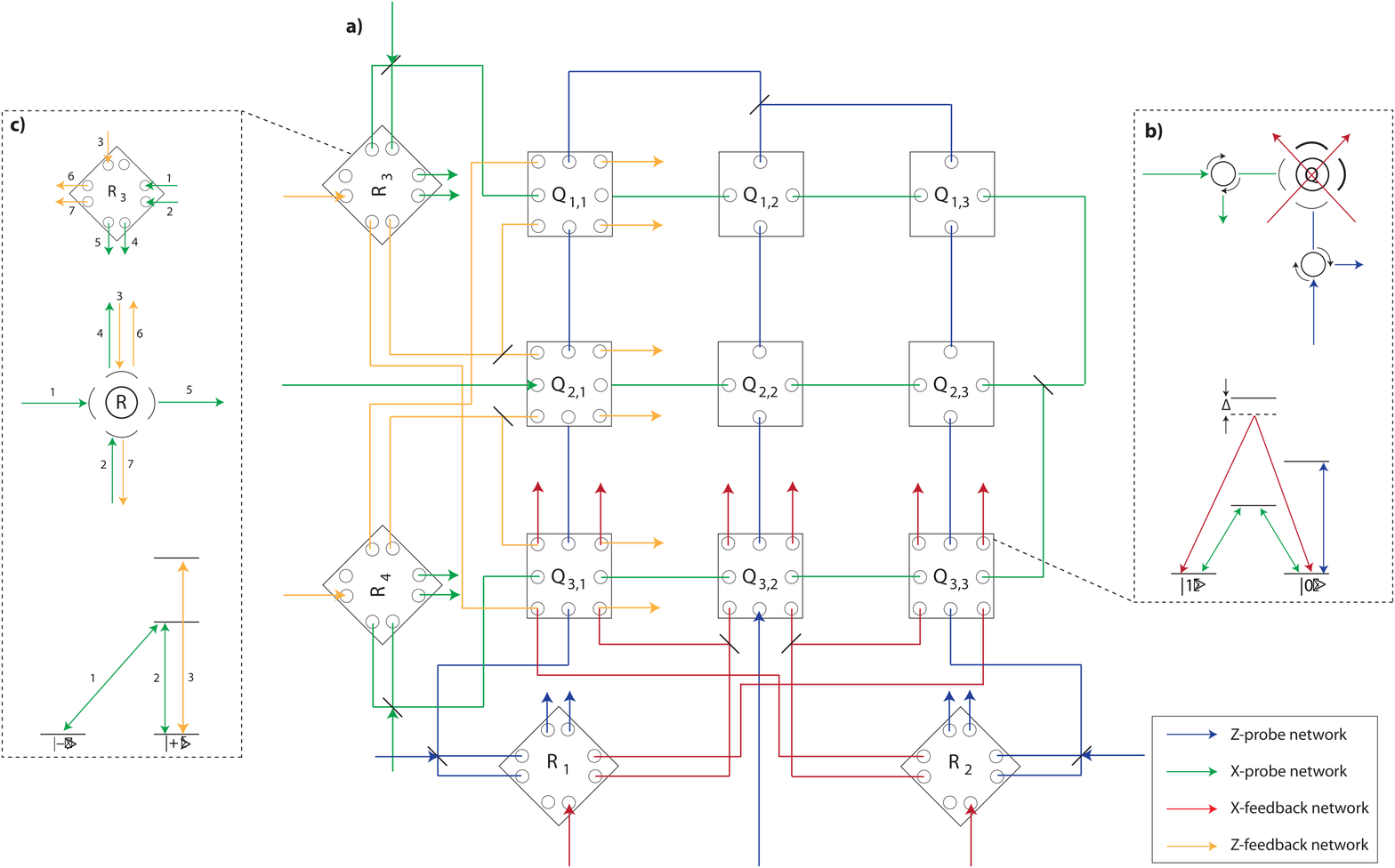}\end{center}
\caption{\label{fig:network}  {\bf a)} Schematic of nanophotonic network capable of implementing the 9 qubit Bacon-Shor QEC code.  CW coherent field inputs that probe the ``Z'' and ``X'' syndromes of the memory qubits, $Q_{i,j}$, enter from the middle of the bottom and left-hand side, in blue and green, respectively.  After traversing the memory qubits, the phases of these fields represent measurements of the four syndrome generators.  Through interference with four more cw ``local oscillator'' laser inputs on beamsplitters and interaction with four ``relay controller'' qubits, $R_i$, these phases effectively control the relays' internal states.  The relay internal states then direct four ``feedback'' cw inputs towards the memory qubits.  When two red (orange) feedback beams simultaneously illuminate a memory qubit, coherent Pauli-X (-Z) rotations occur until a ``no-error'' syndrome state is recovered, at which point the corrective feedback dynamics automatically shut off.  {\bf b)} \& {\bf c)} Example memory and relay cQED input-output, internal level structure, and coupled atomic transition schematics, adapted from \cite{Kerc10}. } 
\vspace{-0.1in}
\end{figure}

As in \cite{Kerc10}, the memory storage qubits are physically realized by multi-level ``atoms'' with two ground states that represent the spin-up and -down states of an ideal qubit.  When an excited state couples to only one ground state (in some basis) via an electric dipole transition that is degenerate with and strongly coupled to a mode of a single-sided optical resonator, then, in appropriate limits, an on-resonance cw laser beam may scatter off the resonator without dissipation or perturbing the qubit state, but will acquire a $\pi$ phase shift upon reflection if the atom is in its coupled ground state, or no phase shift if in its uncoupled state.  This probe interaction approaches this ideal as the atom-mode and waveguide-mode coupling rates dominate all other dynamical rates (dissipative process rates, in particular) \cite{Kerc09a}, which is the natural limit to consider in a ``small-volume,'' quantum nanophotonic context.  The identification of the physical, multi-level system with the ideal, dissipation-less system is formalized by applying these limits and an adiabatic elimination theorem built to work in our quantum optical framework \cite{BvHS08}.  Whereas in the bit-/phase-flip network \cite{Kerc10}, only one type of probe interaction was required per qubit, in this more powerful network, each memory qubit is equipped with the necessary atomic transitions and single-sided resonator modes required to probe the atomic ground state in a Pauli-Z basis with one external probe field and in a Pauli-X basis in another.  Similarly, the controller cQED devices approach their ideal functionality in ``small-volume'' limits equivalent to those of the memory devices \cite{Mabu09b}.  In these limits, each operates as an optical set-reset relay that directs the optical power input of one channel out of one of two output channels depending on the relay's internal state.  This binary internal state is in turn controlled by the presence or absence of optical power entering two additional optical inputs.

Using these components and the intuition gained in \cite{Kerc10}, it's in fact immediately clear how one could design a self-correcting nanophotonic network that emulates the 9 qubit Bacon-Shor QEC code \cite{Baco06}.  In addition, one may also immediately write down the full master equation that governs the internal dynamics of all memory and controller devices of this network without having to resort to a quantum circuit model, suggesting the general applicability of the approach (a laborious, but straightforward network calculation confirms the master equation; see below).  In figure \ref{fig:network} we sketch the nanophotonic design.  Two probe laser inputs first travel down the middle row and column of a $3\times3$ grid of memory qubits.  At the end of the middle row or column, both probes are split on a beamsplitter with each of the now four probe beams traveling back along the remaining rows and columns, respectively.  The frequency and polarization of the laser probe traversing the rows (columns) is such that upon reflection from each memory cQED device, the probe picks up either a $\pi$ or 0 phase shift according to the atomic ground state in a Pauli-X (Pauli-Z) basis.  Thus after traversing the memory qubits, the phases of the four probe lasers become entangled with the memory qubits via controlled-NOT interactions (in a qubit Z- or X-basis) -- essentially as four ancilla qubits would be to extract the error syndrome in a measurement-based QEC scheme \cite{Gott09,QCQI,Alif07}.  By interfering with four local oscillator lasers, the phases of the four probes set the internal state of four relays, which thus play the role of the classical register in a measurement-based scheme; the four relay states together represent the quantum memory's error syndrome \cite{Gott09,QCQI}.  As in \cite{Kerc10}, these relays control the routing of four more input ``feedback'' lasers that perform corrective, unitary rotations on the ground state of the memory qubits via Raman interactions: when a single qubit is simultaneously illuminated by feedback beams emitted by both ``Z-syndrome'' (``X-syndrome'') relays, it undergoes coherent Pauli-X (Pauli-Z) rotations between its ground states (for simplicity in this much larger network, we assume Stark-shift compensation mechanisms are in place \cite{Kerc10}) until the memory qubits -- followed by the controllers -- recover their ``no-error'' state, and automatically shut off the corrective feedback.  Although errors may occur to any of memory qubits, corrective feedback is applied to only a small subset of the memory qubits, as in figure \ref{fig:network}, due to the subsystem structure of the code \cite{Krib05,Baco06} (see \ref{BS9}).  Moreover, as the internal states of the relays are continually reenforced by the cw probes, errors in the relays' atomic states should be self-corrected by the network.  

\begin{figure}[tb!]
\begin{center}\includegraphics[width=0.60\textwidth]{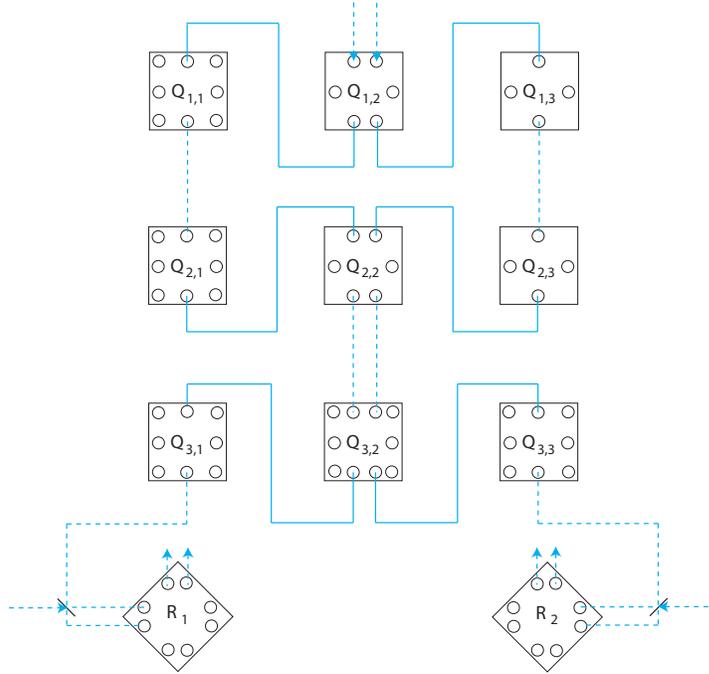}\end{center}
\caption{\label{fig:zigzag}  Schematic of the ``zig-zag'' configuration of the Z-probe network only.  Due to the subsystem structure of the code, dotted connections are capable of suffering calibrated waveguide loss without disruption to the logical information stored in the memory qubits (see \ref{BS9}).  The ``zig-zag'' configuration for the X-probe network follows analogously, essentially rotated by 90$^\circ$.} 
\vspace{-0.1in}
\end{figure}

The subsystem structure of the Bacon-Shor code \cite{Krib05,Baco06} may be further leveraged in a slightly modified network configuration.  Although doing so would require a significantly more intricate waveguide network and/or cQED devices, instead of threading the probe fields directly along rows and columns, a zig-zag configuration, as in figure \ref{fig:zigzag}, should aid network robustness against optical waveguide loss.  Whereas photon loss almost anywhere in error-probe waveguide depicted in figure \ref{fig:network} effectively causes network-induced ``errors'' (which sometimes can be ameliorated by the correction network), logical fidelity in the ``zig-zag'' network should be immune to any waveguide loss in half of the probe network connections \cite{Alif07} (see \ref{BS9}).  We will consider these losses and other critical robustness concerns in depth a later publication, however, and so for our purposes here, both networks represented in figures \ref{fig:network} and \ref{fig:zigzag} have equivalent dynamics. 

The quantum circuit model approach to deriving the master equation dynamics of the network depicted in figure \ref{fig:network} proceeds by appropriately ``connecting'' individual quantum optical device models in series and parallel along freely-propagating optical field modes.  In our quantum stochastic differential equation (QSDE) formalism \cite{HP84}, each device is characterized by a triple $(S,L,H)$, where $H$ is the Hamiltonian of the device's internal degrees of freedom;  $L$ is a vector of coupling operators, each element of which causes these internal variables interact with a particular free field mode; and $S$ is a scattering matrix for the input and output field modes, whose coefficients can in general be operators on the internal state space of the device \cite{Gough09}.  Composite devices may be constructed by feeding the field output(s) of one device directly into another using the ``series product'' operation, $\triangleleft$: $(S_{21},L_{21},H_{21}) = (S_{2},L_{2},H_{2})\triangleleft(S_{1},L_{1},H_{1})$ where $S_{21} = S_2S_1,\ L_{21} = L_2+S_2L_1$, and $ H_{21} = H_1+H_2+$Im$[L_2^\dag S_2L_1]$, representing a system in which the output(s) of device 1 feed(s) into 2.  Devices may also be connected in parallel using the ``concatenation product,'' $\boxplus$, by which, for example, two single-input/-output devices may be aggregated into one dual-input/-output device: $(diag\{S_2,S_1\},\{L_2,L_1\},H_2+H_1) = (S_{2},L_{2},H_{2})\boxplus(S_{1},L_{1},H_{1})$.  The catalog of the quantum devices  in our network (each with an associated triple) includes \cite{Kerc10} the $n$-field identity element, $I_n$; cw laser inputs with amplitude $\gamma$, $W^\gamma$; 50/50 beamsplitters, $B$; the Z-probe (X-probe) aspect of each memory qubit, $Q_{i,j}^Z$ ($Q_{i,j}^X$); the set-reset control of the relays, $R_i^c$; the feedback routing control by the relays, $R_i^f$; and the individual drive-branches of each Raman-X (-Z) interaction in of the memory qubits, $Q_{i,j}^{Xk}$ ($Q_{i,j}^{Zk}$) $k\in\mathbb{Z}_2$.  With these in hand, the appropriate network model may be essentially read off from diagrams like figure \ref{fig:network}.  For example, the entire, closed loop QSDE model for our 9 qubit Bacon-Shor network may be constructed using these individual elements by
\begin{eqnarray}\label{eq:network}
\fl G_p^Z = \left(R_1^c\boxplus R_2^c\right)\triangleleft\left(B\boxplus B\right)\triangleleft\left(\left(Q_{3,1}^Z\triangleleft Q_{2,1}^Z\triangleleft Q_{1,1}^Z\right)\boxplus I_1\boxplus\left(Q_{3,3}^Z\triangleleft Q_{2,3}^Z\triangleleft Q_{1,3}^Z\right)\right)\triangleleft\nonumber\\
\left(\left(B\boxplus_2 I_1\right)\boxplus I_1\right)\triangleleft\left(\left(Q_{1,2}^Z\triangleleft Q_{2,2}^Z\triangleleft Q_{3,2}^Z\triangleleft W^{\sqrt2\alpha}\right)\boxplus W^\alpha\boxplus I_1\boxplus W^\alpha\right)\nonumber\\
\fl G_{f1} = \left(Q_{3,3}^{X1}\boxplus Q_{3,1}^{X1}\boxplus Q_{3,2}^{X1}\right)\triangleleft\left(I_1\boxplus B\right)\triangleleft\left(\left(R_1^f\triangleleft\left(W^\beta\boxplus I_1\right)\right)\boxplus I_1\right)\nonumber\\
\fl G_{f2} = \left(Q_{3,1}^{X1}\boxplus Q_{3,3}^{X1}\boxplus Q_{3,2}^{X1}\right)\triangleleft\left(I_1\boxplus B\right)\triangleleft\left(\left(R_2^f\triangleleft\left(W^\beta\boxplus I_1\right)\right)\boxplus I_1\right)\nonumber\\
\fl G_p^X = \left(R_3^c\boxplus R_4^c\right)\triangleleft\left(B\boxplus B\right)\triangleleft\left(\left(Q_{1,1}^X\triangleleft Q_{1,2}^X\triangleleft Q_{1,3}^X\right)\boxplus I_1\boxplus\left(Q_{3,1}^X\triangleleft Q_{3,2}^X\triangleleft Q_{3,3}^X\right)\right)\triangleleft\nonumber\\
\left(\left(B\boxplus_2 I_1\right)\boxplus I_1\right)\triangleleft\left(\left(Q_{2,3}^X\triangleleft Q_{2,2}^X\triangleleft Q_{2,1}^X\triangleleft W^{\sqrt2\alpha}\right)\boxplus W^\alpha\boxplus I_1\boxplus W^\alpha\right)\nonumber\\
\fl G_{f3} = \left(Q_{3,1}^{Z1}\boxplus Q_{1,1}^{Z1}\boxplus Q_{2,1}^{Z1}\right)\triangleleft\left(I_1\boxplus B\right)\triangleleft\left(\left(R_3^f\triangleleft\left(W^\beta\boxplus I_1\right)\right)\boxplus I_1\right)\nonumber\\
\fl G_{f4} = \left(Q_{1,1}^{Z1}\boxplus Q_{3,1}^{Z1}\boxplus Q_{2,1}^{Z1}\right)\triangleleft\left(I_1\boxplus B\right)\triangleleft\left(\left(R_4^f\triangleleft\left(W^\beta\boxplus I_1\right)\right)\boxplus I_1\right)\nonumber\\
\fl G_{network} = G_p^Z\boxplus G_{f1}\boxplus G_{f2}\boxplus G_p^X\boxplus G_{f3}\boxplus G_{f4}
\end{eqnarray}
where $G_p^Z$ ($G_p^X$) describes the ``Z'' (``X'') syndrome extraction network, $G_{fi}$ describes the feedback network controlled by relay $i$, and $\boxplus_2$ is a ``padding operator'' related to the concatenation product \cite{Kerc10} (necessary for proper field indexing), and $G_{network}$ represents the $(S,L,H)$ triple of the entire system, describing all but the error dynamics.  While the symbolic calculation of the analogous $G_{network}$ was done by hand in \cite{Kerc10}, this tedious calculation may be automated using simple symbolic matrix manipulation scripts in programs such as Mathematica, freeing a quantum network designer to easily edit and adjust systems essentially at the schematic level.  As in \cite{Kerc10}, the final step in the calculation, which yields a relatively low-dimensional total network $G_{network}$, is the application of an algorithmic adiabatic elimination procedure \cite{BvHS08} that restricts the Raman interaction dynamics to the ground states of the memory qubits only.   

Including arbitrary single-qubit dephasing of each memory qubit to $G_{network}$ through simple concatenation products, the closed loop dynamics of $\rho_t$, the density matrix for all 9 memory and 4 relays qubits (physically representing the entire system's dynamics after tracing over the field degrees of freedom), is read off from the entire system's $(S,L,H)$ triple \cite{Kerc10,OSQO}, giving:
\begin{eqnarray}\label{eq:ME}
\dot{\rho}_t &=&-i[H,\rho_t] + \sum_{j=1}^{35}\left( L_j\rho_tL_j^* - \frac{1}{2}\{L_j^*L_j,\rho_t\}\right),\quad\mathrm{with}\nonumber\\
H&=&\Omega\biggl(\sqrt{2}X_{3,1}\Pi^-_{1}\Pi^+_{2}+X_{3,2}\Pi^-_{1}\Pi^-_{2}-\sqrt{2}X_{3,3}\Pi^+_{1}\Pi^-_{2}+\nonumber\\
&&\sqrt{2}Z_{1,1}\Pi^-_{3}\Pi^+_{4}+Z_{2,1}\Pi^-_{3}\Pi^-_{4}-\sqrt{2}Z_{3,1}\Pi^+_{3}\Pi^-_{4}\biggr),\nonumber\\
L_1 &=& \frac{\alpha}{\sqrt2}\left(\Pi^-_{1}O^Z_{1\&2}+\sigma^{+-}_{1}E^Z_{1\&2}\right),\quad L_2  = \frac{\alpha}{\sqrt2}\left(-\sigma^{-+}_{1}O^Z_{1\&2}-\Pi^+_{1}E^Z_{1\&2}\right),\nonumber\\
L_3 &=& \frac{\alpha}{\sqrt2}\left(\Pi^-_{2}O^Z_{3\&2}+\sigma^{+-}_{2}E^Z_{3\&2}\right),\quad L_4  = \frac{\alpha}{\sqrt2}\left(-\sigma^{-+}_{2}O^Z_{3\&2}-\Pi^+_{2}E^Z_{3\&2}\right),\nonumber\\
L_5 &=& \frac{\alpha}{\sqrt2}\left(\Pi^-_{3}O^X_{1\&2}+\sigma^{+-}_{3}E^X_{1\&2}\right),\quad L_6  = \frac{\alpha}{\sqrt2}\left(-\sigma^{-+}_{3}O^X_{1\&2}-\Pi^+_{3}E^X_{1\&2}\right),\nonumber\\
L_7 &=& \frac{\alpha}{\sqrt2}\left(\Pi^-_{4}O^X_{3\&2}+\sigma^{+-}_{4}E^X_{3\&2}\right),\quad L_8  = \frac{\alpha}{\sqrt2}\left(-\sigma^{-+}_{4}O^X_{3\&2}-\Pi^+_{4}E^X_{3\&2}\right),\nonumber\\
L_{9-17} &=& \sqrt{\Gamma}X_{i,j},\quad L_{18-26} = \sqrt{\Gamma}Z_{i,j}, \quad L_{26-35} = \sqrt{\Gamma}Y_{i,j}\quad i,j\in\mathbb{Z}_3,
\end{eqnarray}
where $\{X_{i,j},Z_{i,j}\}$ are the Pauli-$\{$X,Z$\}$ operators on the memory qubit in row $i$ and column $j$, $\Pi^\pm_i$ is the projector onto state $\vert\pm\rangle$ of relay $i$, $\sigma_i^{\pm\mp}=\vert\pm\rangle\langle\mp\vert$ for relay $i$, and $\{-O^Z_{i\&j},E^Z_{i\&j},-O^X_{i\&j},O^X_{i\&j},\}=1+\{-Z_{\ast,i}Z_{\ast,j},Z_{\ast,i}Z_{\ast,j},-X_{i,\ast}X_{j,\ast},X_{i,\ast}X_{j,\ast}\}$ with index $\ast$ signifying the operator product of all operators acting on some row or column.  $\Gamma$ is the mean error rate of each type of single-qubit error, $\alpha$ is the coherent amplitude of the (eventual) four probe lasers (where $\vert\alpha\vert^2$ has units photons time$^{-1}$), and $\Omega$ is proportional to the optical power in each feedback beam.  Comparison with the master equation derived in \cite{Kerc10}, reveals that these dynamics for the 9 qubit QEC subsystem code appear as an almost trivial expansion of that simple bit-/phase-flip network.

\begin{figure}[tb!]
\begin{center}\includegraphics[width=0.80\textwidth]{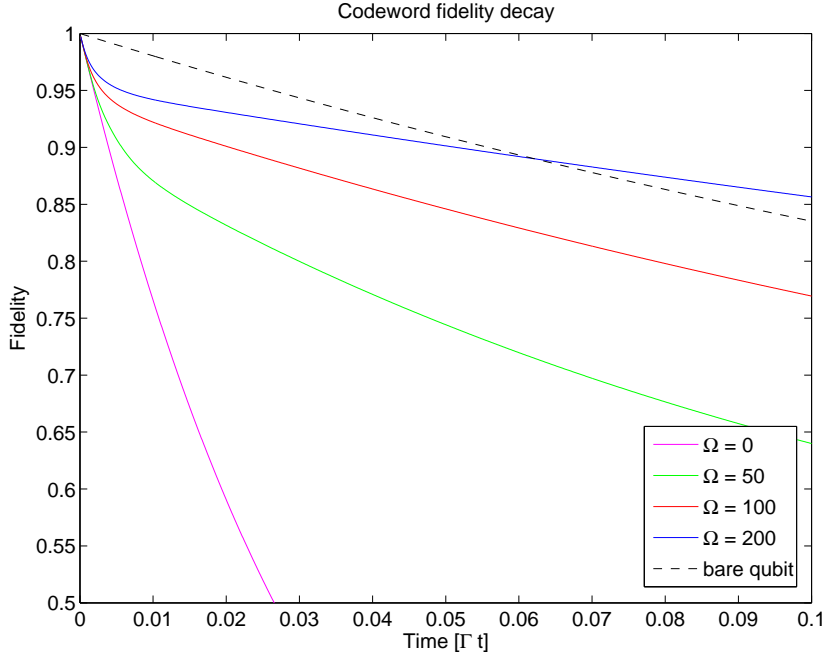}\end{center}
\caption{\label{fig:fid}  Decay of the logical codeword fidelity (see \ref{BS9}) in time for several feedback parameters.  In the simulation, single-memory qubit X-, Y- and Z-errors occur each at average rate $\Gamma = .1$ per qubit.  While the amplitude of the probes, are fixed at $\alpha=\Omega/8$, the feedback ``strength''  $\Omega$ is varied from 0 (no feedback) to 200.  For large feedback strengths, the logical storage fidelity is eventually superior to that of a single, bare qubit suffering arbitrary dephasing.} 
\vspace{-0.1in}
\end{figure}

Despite our idealized description of all cQED systems as simple qubits, Eq. \ref{eq:ME} still represents complex dynamics on a (sparse) density matrix with dimensions $2^{9+4}\times2^{9+4}$.  Due to the size of the problem, multi-threaded numerical integration of Eq. \ref{eq:ME} was carried out on a multicore computational server using BLAS routines with OpenMP.  Figure \ref{fig:fid} depicts the time-evolution of the fidelity of the logical information stored in the system, initialized with the logical qubit in the +1 $Y_L$ eigenstate, the four gauge qubits in +1 $Z_{Gi}$ eigenstates (see \ref{BS9}), and the four relays each in their ``no error'' $\vert+\rangle$ state.  The bitwise-error rate was fixed at $\Gamma = .1$, the probe strength kept at $\alpha = \Omega/8$, and the ``feedback strength'' $\Omega$ varied from 0 to 200.  Initially, all fidelity curves drop steeply, representing a network ``latency'' before both probe and feedback networks begin to correct errors.  Thereafter, fidelity still drops due to the finite probability that multiple errors irrecoverably accumulate before correction, but at a retarded rate that decreases with increasing feedback strength.  For $\Omega > 100$, despite the initially steep fidelity loss, the decay is slow enough to eventually achieve superior storage fidelity than a single, uncorrectable qubit suffering X-, Y-, and Z-errors at the same bit-wise rate.

\begin{figure}[tb!]
\begin{center}\includegraphics[width=0.80\textwidth]{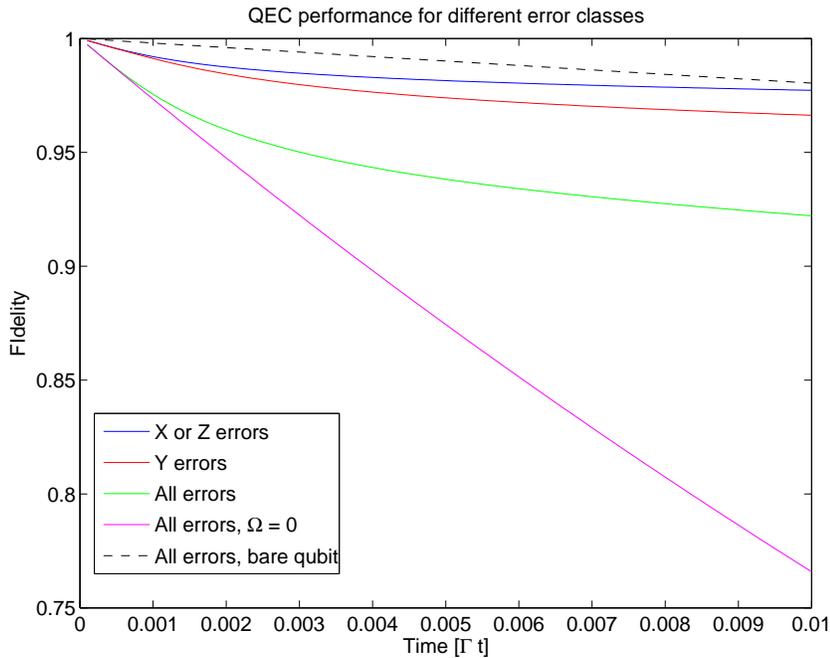}\end{center}
\caption{\label{fig:DErr}  Comparison of network storage fidelity between systems suffering different types of errors.  All solid curves are simulated with $\{\Gamma,\alpha,\Omega\} = \{.1,100/8,100\}$, except for the magenta curve, which has no feedback, as indicated.  The blue curve represents the storage fidelity when only X- or only Z-errors occur to each memory qubit.  The only slightly lower red curve comes from a system suffering only Y-errors, which require coordination between all aspects of the network.  The green curve represents a system suffering all three types of error, each at the same bit-wise rate, and has roughy the same fidelity decay as a sum of the decays from each type of error acting alone.  The dashed black line again represents the logical fidelity of a single, bare qubit suffering all three types of errors at mean rate $\Gamma = .1$ each.} 
\vspace{-0.1in}
\end{figure}

We may begin to break down the dynamics of this particular network by calculating its performance in response to different types of errors.  In figure \ref{fig:DErr}, we plot the fidelity decay in time with $\{\Gamma,\alpha,\Omega\} = \{.1,100/8,100\}$, but with the memory qubits suffering only single-qubit X-, Z-, or Y-errors.  Systems suffering only X- or Z-errors require only ``half'' of the feedback network to function well and achieve the best storage fidelity.  The system responds slightly worse to Y-errors.  This is expected as Y-flip errors (i.e. simultaneous bit-flip {\it and} phase-flip errors on a single qubit) requires the two halves of the correction network to work in concert to correct the error.  That such coordination occurs without too much additional fidelity loss suggests that the apparatus is fairly unbiased towards the types of errors it corrects.  Similarly, when all three errors are acting simultaneously, each at mean rate $\Gamma = .1$ per qubit, the additional fidelity loss scales roughly linearly with the {\it total} error rate in this strong feedback regime, as one would expect for an efficiently operating system.

In conclusion, we have demonstrated that the concept of autonomous nanophotonic quantum circuits is sufficiently flexible to be straightforwardly applied in designing robust and naturally fault-tolerant quantum memories that emulate measurement-based approaches like the 9 qubit Bacon-Shor QEC code.  Modeled with continuous-time dynamics, our models are built upon true {\it quantum circuit diagrams} like figure \ref{fig:network}, in which components represent physical devices (not discrete operations) and connections represent physical optical fields (not just time-ordered process flows).  Future directions in our work will focus on developing freely-available software packages that will allow users to design, analyze, and simulate similar networks with configurable connections and fully-specifiable Hamiltonian devices truly at this schematic level, with only minimal expertise in the underlying formalisms.  On-going work directly related to these memory networks include analyzing robustness to mis-tuned parameters and optical losses, formulating largely automated quantum memory registry ``read-in/-out'' procedures compatible with these networks, and considering protection schemes beyond simple emulation of well-known quantum stabilizer codes.

\section{Acknowledgements}
This work has been supported by the ARO (W911NF-08-1-0427) and by the NSF (PHY-1005386).  DSP and HC acknowledge the support of Stanford Graduate Fellowships.  The authors would like to thank Hendra Nurdin for useful discussions and critical reading of the manuscript.

\appendix
\section{The 9 qubit Bacon-Shor QEC code}\label{BS9}
This appendix summarizes the measurement-based 9 qubit Bacon-Shor QEC \cite{Baco06,Alif07} subsystem code our autonomous photonic network emulates.  As in the main article, the 9 qubits are arranged in a $3\times3$ grid.  In this configuration, a non-Abelian operator group is the set of Pauli operators generated by pairs of X operators acting on adjacent-row qubits and pairs of Z operators acting on adjacent-column qubits
\begin{equation}
\mathcal{T} = \langle X_{i,j}X_{i+1,j},Z_{j,i}Z_{j,i+1}\vert i\in\mathbb{Z}_2,j\in\mathbb{Z}_3\rangle,
\end{equation}
where $X_{i,j}$ and $Z_{i,j}$ act as Pauli operators on the qubit in row $i$, column $j$ and as the identity on all others.  A subgroup $\mathcal{S}\subset\mathcal{T}$ serves as the stabilizer \cite{Gott09,QCQI} for the code and is generated by X operators acting on all qubits in two adjacent rows and Z operators acting on all qubits in two adjacent columns
\begin{equation}
\mathcal{S} = \langle X_{i,*}X_{i+1,*},Z_{*,i}Z_{*,i+1}\vert i\in\mathbb{Z}_2\rangle.
\end{equation}
As $\mathcal{S}$ is an Abelian group of observables, the $\pm1$ eigenvalues of the generators of $\mathcal{S}$ may be used to label subspaces of the $\mathcal{H} = (\mathbb{C}^2)^{\otimes9}$ 9-qubit Hilbert space
\begin{equation}
\mathcal{H} = \bigoplus_{\vec a}\mathcal{H}_{\vec a}
\end{equation} 
where $\vec a$ is string of the four binary eigenvalues of the stabilizer generators and each $\mathcal{H}_{\vec a}$ represents a $2^5$-dimensional space.  Each $\mathcal{H}_{\vec a}$ may be further decomposed into subsystems \cite{Krib05,Baco06,Alif07}
\begin{equation}
\mathcal{H}_{\vec a} = \mathcal{H}_{\vec a}^\mathcal{T}\otimes\mathcal{H}_{\vec a}^\mathcal{L},
\end{equation}
where elements from $\mathcal{T}$ (which commute with $\mathcal{S}$) act non-trivially on some $\mathcal{H}_{\vec a}^\mathcal{T}$, and as the identity on $\mathcal{H}_{\vec a}^\mathcal{L}$, encoding four logical qubits, modulo $\mathcal{S}$.  This action leaves the $\mathcal{H}_{\vec a}^\mathcal{L}$ as single-qubit subsystems, in one of which our logical qubit codeword is stored.  We can choose this logical qubit to be the one encoded by the operators $X_L = X_{1,\ast}$ and $Z_L = Z_{\ast,1}$ in the $\vec a = \{+1,+1,+1,+1\}$ stabilizer subspace.  The other four logical qubits that live in this subspace's $\mathcal{H}_{\vec a=1^{\otimes4}}^\mathcal{T}$ subsystem may be specified by four commuting pairs of anticommuting operators in $\mathcal{T}$: $\{X_{Gi},\ Z_{Gi}\vert i\in\mathbb{Z}_4\}$.  The logical qubits that live in $\mathcal{H}_{\vec a=1^{\otimes4}}^\mathcal{T}$ are deemed ``gauge'' qubits that have no information encoded in them by the user.  The QEC procedure that protects the single logical qubit codeword in $\mathcal{H}_{\vec a=1^{\otimes4}}^\mathcal{L}$ may perturb the state of these gauge qubits without consequence.

In the measurement-based QEC code, the encoded codeword may be rendered impervious to arbitrary single-qubit rotations of any of the nine qubits, although the state of gauge qubits may be depolarized.  Measurement of the four stabilizer generators ``discretizes'' any partial qubit rotations, projecting the system state into some error space $\mathcal{H}_{\vec a}$, with $\vec a =  \{+1,+1,+1,+1\}$ being the ``no error'' space.  The results of these syndrome measurements localize Pauli-X errors down to the column in which they occurred, Pauli-Z down to the row, and Pauli-Y down to the qubit.  Although the precise location of the error is not known generally, codeword recovery is achieved by applying a Pauli-X (-Z) to {\it any} qubit in the same column (row) as a detected bit-flip (phase-flip) error.  For example, if a partial Pauli-Y rotation occurs to qubit $Q_{2,2}$ (the middle qubit in the $3\times3$ grid), syndrome measurement may project the system into the $\vec a =  \{-1,-1,-1,-1\}$ subspace, digitizing and indicating the error.  The encoded codeword may be recovered by then applying the operator $X_{3,2}Z_{2,1}$, as the net, recovery-error operator, $X_{3,2}Z_{2,1}Y_{2,2}$, commutes with $\mathcal{S}$, $X_L$ and $Z_L$.  However, $X_{3,2}Z_{2,1}Y_{2,2}\in\mathcal{T}$ and thus the recovery operation disturbs the state of the logical gauge qubits.  Moreover, the error syndrome need not be acquired by directly measuring the 6-body operator generators \cite{Alif07}.  Separably measuring the 2-body generators of $\mathcal{T}$ yields the same syndrome information as the generators of $\mathcal{S}$ and although such measurements perturb the logical gauge qubits, they commute with $X_L$ and $Z_L$ and thus preserve the codeword.  This remarkable fact is the motivation behind the ``zig-zag'' network in figure \ref{fig:zigzag}, which should be more robust to waveguide loss along much of the probing network. 

The fidelity measure used in figure \ref{fig:fid} was calculated from $\rho_t$ at each time $t$ as Tr($F\rho_t$), where $F$ is the ``fidelity operator'' constructed by taking the initial state $\rho_0$, completely depolarizing all four ``gauge'' and four relay qubits, and multiplying by $2^{4+4}$.  Thus, the expectation of $F$ compares the state at time $t$ to that at 0, tracing over all but the logical qubit in the ``no-error'' syndrome subspace. 



\section*{References}

\begin{thebibliography}{99}
\bibitem{Gott09} D.~Gottesman 2009 quant-ph/0904.2557v1

\bibitem{QCQI} Nielsen~M~A and Chuang~I~L~2000 {\it Quantum computation and quantum information} (Cambridge: Cambridge University Press). 

\bibitem{Kerc10} Kerckhoff~J, Nurdin~H~I, Pavlichin~D~S and Mabuchi H 2010 {\it Phys. Rev. Lett.} {\bf 105} 040502

\bibitem{Mabu09b} Mabuchi~H 2009 {\it Phys.\ Rev.\ A} {\bf 80} 045802 (2009).

\bibitem{Baco06} Bacon~D 2006 {\it Phys.\ Rev.\ A} {\bf 73} 012340

\bibitem{Alif07} Aliferis P and Cross A W 2007 {\it Phys.\ Rev.\ Lett.} {\bf 98} 220502

\bibitem{Gardin92} Gardiner C W, Parkins A S and Zoller P 1992 {\it Phys. Rev. A.} {\bf 46} 4363

\bibitem{OSQO} Carmichael H 1993 {\t An open systems approach to quantum optics} (Berlin: Springer-Verlag)

\bibitem{Gough09} Gough J and James M R 2009 {\it IEEE Trans Automat. Contr.} {\bf 54} 2530

\bibitem{Nurd09a}  Nurdin H I, James M R and Doherty A C 2009 {\it SIAM J. Control Optim.} {\bf 48} (4) 2686-2718

\bibitem{HP84} Hudson R L and Parthasarathy K R 1984 {\it Comm.\ Math.\ Phys.\ } {\bf 93} 301

\bibitem{Kerc09a} Kerckhoff J, Bouten L, Silberfarb A and Mabuchi H 2009 {\it Phys.\ Rev.\ A} {\bf 79} 024305

\bibitem{BvHS08} Bouten L, van~Handel R and Silberfarb A 2008 {\it J. Funct. Anal.} {\bf 254} 3123

\bibitem{Krib05} Kribs D, Laflamme R and Poulin D 2005 {\it Phys.\ Rev.\ Lett.} {\bf 94} 180501
\end{thebibliography}
\end{document}